\documentclass [12 pt]{article}
\def\be{\begin{equation}}
\def\eea{\end{eqnarray}}
\def\bea{\begin{eqnarray}}
\def\ee{\end{equation}}

\usepackage[psamsfonts]{amssymb}
\usepackage{amsmath}
\author{M. Amooshahi$^{1}$ \footnote{amooshahi@sci.ui.ac.ir}
\\ $^{1}$ {\small Faculty of science, University of Isfahan ,Hezar Jarib Ave.,
Isfahan,Iran}}
\title{Spin-$\frac{1}{2}$ particle in an absorbing environment}
\begin{document}
\maketitle
\begin{abstract}
The quantum dynamics of a localized spin-$\frac{1}{2} $ Particle
interacting with an absorbing environment is investigated. The
quantum Langevin-Schr\"{o}dinger equation for spin-$\frac{1}{2} $ is
obtained. The susceptibility function of the environment is
calculated in terms of the coupling function of the spin and the
environment. it is shown that the susceptibility function satisfies
the Kramers-Kronig relations. Spontaneous emission and the shift
frequency of the spin is obtained in terms of the imaginary part of
the  susceptibility function in frequency domain. Some transition
probabilities  between the spin states  are calculated when the
absorbing environment is in the thermal state.
\end{abstract}
\section{Introduction}
There are mainly two approaches to study dissipative quantum
systems. One approach is a phenomenological treatment under the
assumption of nonconservative forces\cite{1,2}. The second approach
is found in the interaction between two systems via an irreversible
energy flow
\cite{3}-\cite{4.1}.\\
  In the frame work of the first approach  in studying
nonconservative systems, it is essential to introduce a time
dependent Hamiltonian which describes the damped motion. Such a
phenomenological approach for the study of dissipative quantum
systems, specially a damped harmonic oscillator, has a rather long
history. Caldirola and Kanai \cite{5,6} adopted a Hamiltonian for a
harmonic oscillator so that the Heisenberg equation of the
oscillator  is identical to the classical equation of a damped
harmonic oscillator. The quantum aspect of this model has been
studied in a great amount of literature. In those studies some
peculiarities of this model and some features of it have appeared to
be ambiguous \cite{7}-\cite{15}. There are significant difficulties
in obtaining the quantum mechanical solutions for the
Caldirola-Kanai Hamiltonian. Quantization with this Hamiltonian
violates the uncertainty relations. That is the uncertainty
relations vanish as time goes to infinity,\cite{16}-\cite{19}. Based
on Caldirola-Kanai Hamiltonian ,  equivalent theories have been
constructed by performing a quantum canonical transformation. Also
 the path integral techniques has been used to calculate the
exact propagators of such theories. The time evolution of a given
initial wave functions have been studied using the obtained
propagators \cite{20}. In the framework of the phenomenological
approach Lopez and Gonzales \cite{21} have taken the external non
conservative forces that has linear and quadratic dependence  with
respect to velocity. They have deduced classical constants of motion
and Hamiltonian for these systems. The eigenvalues of these
constants have been quantized through perturbation theory.\\
In the second approach to study quantum dissipative systems  one
tries to bring about the dissipation as a results of an averaging
over all the coordinates of a bath system. One considers the whole
system as composed of two parts, our main system and the bath system
which interacts with the main system and causes the dissipation of
energy on it\cite{22}-\cite{29}. The macroscopic description of a
quantum particle with dissipation  and moving in an external
potential is formulated in terms of the Langevin- Schr\"{o}dinger
equation  \cite{30,31}
\begin{equation}
m\ddot{\vec{x}}+\int_0^t dt'
\mu(t-t')\dot{\vec{x}}(t')=-\vec{\nabla}v(\vec{x})+\vec{F}_N(t)
\end{equation}
 The coupling with the heat-bath in
microscopic levels correspond two terms in macroscopic description.
A mean force characterized by a memory function $\mu(t)$ and an
operator valued random force $ \vec{F}_N(t)$. These two terms have a
fluctuation-dissipation relation and both are required for a
consistent quantum mechanical description of the particle. In
\cite{31} there are some models for interaction of the main system
with the heat-bath which leads to macroscopic
Langevin-Schr\"{o}dinger equation.\\
In the present paper we use the second approach for a damped
spin-$\frac{1}{2}$ particle embedded in a uniform magnetic field. In
section 2 the absorbing environment of the spin is modeled by a
continuous set of harmonic oscillators and quantum dynamics of the
spin is investigated. In section 3 the spontaneous emission and the
shift frequency of the spin are calculated. In section 4 some
transition probabilities between the eigenstates of the spin are
computed when the state of the environment is thermal state.
\section{Quantum dynamic}
Suppose a localized spin-$\frac{1}{2}$ particle is in a constant
magnetic field $\vec{B}=B_0\hat{n}$ along a unit vector $\hat{n}$
and interacts with an absorbing environment. Then the total
Hamiltonian can be written as three parts
\begin{equation}\label {s1}
H=H_s+H_{Bs}+H_B
\end{equation}
$H_s$ is the Hamiltonian of the spin-$\frac{1}{2}$ particle in the
constant magnetic field
\begin{equation}\label {s2}
H_s=\omega_0\vec{S}\cdot\hat{n}
\end{equation}
where $\omega_0=\frac{|e|B_0}{mc}$ and $e,m$ are the charge and mass
of the particle and $c$ is the speed of light. $ H_B$ is the
Hamiltonian of the absorbing environment containing  a continuum set
of harmonic oscillators
\begin{equation}\label {s3}
H_B=\sum_{\lambda=1}^3\int d^3k  \hbar\omega_{\vec{k}}
a_{\vec{k}\lambda}^\dag a_{\vec{k}\lambda}
\end{equation}
where $\omega_{\vec{k}}=c|\vec{k}|$ and $a_{\vec{k}\lambda} ,
a_{\vec{k}\lambda}^\dag$ are annihilation and creation operators of
the absorbing environment, respectively  and satisfy the commutation
relations
\begin{equation}\label {s4}
[a_{\vec{k}\lambda},
a_{\vec{k'}\lambda'}^\dag]=\delta_{\lambda\lambda'}\delta(\vec{k}-\vec{k'}).
\end{equation}
In (\ref{s1}) $ H_{sB}$ is the interaction part of the spin and the
environment and can be proposed as
\begin{equation}\label {s5}
H_{sB}=-\vec{S}\cdot\vec{R}
\end{equation}
where $\vec{R}(t)$ in the Heisenberg picture is as follows
\begin{equation}\label {s6}
\vec{R}(t)=\sum_{\lambda=1}^3\int d^3k[
f(\omega_{\vec{k}})a_{\vec{k}\lambda}(t)+f^{\star}(\omega_{\vec{k}})a_{\vec{k}\lambda}^\dag(t)]\hat{e}_{\vec{k}\lambda}
\end{equation}
In this relation $\hat{e}_{\vec{k}\lambda}\  \lambda=1,2,3$ are
three orthogonal unit vectors and  $f(\omega_{\vec{k}})$ is called
the coupling function  between the spin-$\frac{1}{2}$ and the
environment and has a  crucial role in this theory.\\
It can be easily shown that the equation of motion of the spin in
the Heisenberg picture is obtained as
\begin{equation}\label {s6.1}
\dot{\vec{S}}=\frac{i}{\hbar}[H ,
\vec{S}]\Rightarrow\dot{\vec{S}}=\omega(
\hat{n}\times\vec{S})-\vec{R}\times\vec{S}
\end{equation}
where the commutation relations $[S_i,
S_j]=i\hbar\varepsilon_{ijk}S_k$ has been used. Also the Heisenberg
equation for the annihilation operators $a_{\vec{k}\lambda}$ are
\begin{equation}\label {s7}
\dot{a}_{\vec{k}\lambda}(t)=-i\omega_{\vec{k}}
a_{\vec{k}\lambda}(t)+\frac{i}{\hbar} f^\star(\omega_{\vec{k}})
 \vec{S}(t)\cdot\hat{e}_{\vec{k}\lambda}
\end{equation}
This equation can be solved formally as
\begin{equation}\label {s8}
a_{\vec{k}\lambda}(t)=a_{\vec{k}\lambda}(0)e^{-i\omega_{\vec{k}}
t}+\frac{i}{\hbar} f^\star(\omega_{\vec{k}})
\hat{e}_{\vec{k}\lambda}\cdot\int_0^t dt'
e^{-i\omega_{\vec{k}}(t-t')} \vec{S}(t')
\end{equation}
substituting $a_{\vec{k}\lambda}(t)$ from (\ref{s8})  in (\ref{s6})
and doing some calculations give us the time dependence of $
\vec{R}(t)$ as
\begin{equation}\label {s9}
\vec{R}(t)=\vec{R}_N(t)+\int_0^{|t|} dt'\chi(|t|-t') \vec{S}(\pm t')
\end{equation}
where the upper (lower) sign is for $t>0 (t<0)$ and $\chi(t)$ is the
susceptibility tensor of the absorbing environment and is obtained
as
\begin{equation}\label {s10}
\chi(t)=\frac{8\pi}{\hbar c^3}\int_0^\infty d\omega \omega^2
|f(\omega)|^2\sin \omega t \Theta(t)
\end{equation}
where $ \Theta(t)$ is the step function. The equation (\ref{s9}) may
be interpreted the constitutive relation or response equation of the
environment. One can be shown that the susceptibility tensor $\chi$
satisfies the Kromers-Kronig relations\cite{31.1}
\begin{eqnarray}\label {s11}
&&Re [\tilde{\chi}(\omega)]=\frac{2}{\pi}P\int_0^\infty d\omega'
\frac{\omega' I
m[\bar{\chi}(\omega')}{\omega'^2-\omega^2}\nonumber\\
&& I m[\tilde{\chi}(\omega)]=-\frac{2\omega}{\pi}P\int_0^\infty
d\omega' \frac{Re[\bar{\chi}(\omega')}{\omega'^2-\omega^2}
\end{eqnarray}
where the symbol $P$ denote the cauchy principal  value of the
integrals and
\begin{equation}\label {s11.1}
\tilde{\chi}(\omega)=\int_0^\infty dt \chi(t)e^{i\omega t}
\end{equation}
 is Fourier transformation of $\chi(t)$. In (\ref{s9}) $ \vec{R}_N(t)$ is a noise field and is obtained in terms
of the annihilation and creation operators of the environment at
$t=0$ as
\begin{equation}\label {s12}
\vec{R}_N(t)=\sum_{\lambda=1}^3\int d^3k[
f(\omega_{\vec{k}})a_{\vec{k}\lambda}(0)e^{-i\omega_{\vec{k}}t}+f^{\star}(\omega_{\vec{k}})a_{\vec{k}\lambda}^\dag(0)
e^{i\omega_{\vec{k}}t}]\hat{e}_{\vec{k}\lambda}
\end{equation}
Finally insertion $\vec{R}(t)$ from (\ref{s9}) into equation
(\ref{s6.1}) leads to the equation of motion of the spin as
\begin{equation}\label {s13}
\dot{\vec{S}}-\omega(\hat{n}\times\vec{S})+\int_0^{|t|}dt'\chi(|t|-t')(\vec{S}(\pm
t')\times\vec{S}(t))=-\vec{R}_N(t)\times\vec{S}(t)
\end{equation}
 which can be interpreted as the Langevin -Schr\"{o}dinger equation for
the spin in an absorbing environment.
\section{Spontaneous emission}
In this section the spontaneous decay rate of an initially exited
spin-$\frac{1}{2}$ particle is calculated. For simplicity, we assume
the spin is in a constant magnetic field along the $z$ axis.
Therefore the Hamiltonian (\ref{s1}) can be written as
\begin{eqnarray}\label {s14}
&&H=H_0+H'\nonumber\\
 && H_0=\omega_0 S_z+\sum_{\lambda=1}^3\int d^3k  \hbar\omega_{\vec{k}}
a_{\vec{k}\lambda}^\dag a_{\vec{k}\lambda}\nonumber\\
&&H'=-\vec{S}\cdot\vec{R}
\end{eqnarray}
To study the spontaneous emission  the interaction picture is used
and we apply the Weisskopf-Wigner approximation\cite{32}.The
interaction Hamiltonian $H'$ in interaction picture is as follows
\begin{eqnarray}\label {s15}
&&H'_I(t)= -\sum_{\lambda=1}^3\int d^3k
f(\omega_{\vec{k}})e^{i(\omega_0-\omega_{\vec{k}})t}(\frac{e_{\vec{k}\lambda
x}}{2}+\frac{e_{\vec{k}\lambda
y}}{2i})a_{\vec{k}\lambda}(0)S_+(0)\nonumber\\
&& -\sum_{\lambda=1}^3\int d^3k
f^\star(\omega_{\vec{k}})e^{i(\omega_{\vec{k}}-\omega_0)t}(\frac{e_{\vec{k}\lambda
x}}{2}-\frac{e_{\vec{k}\lambda
y}}{2i})a^\dag_{\vec{k}\lambda}(0)S_-(0)\nonumber\\
\end{eqnarray}
where the rotating -wave approximation \cite{32} has been used and
$S_+=S_x+iS_y\ , S_-=S_x-iS_y$. In the framework of the
Weisskopf-Wigner theory the wave function of the total system in
interaction picture is written as
\begin{equation}\label {s16}
|\psi(t)\rangle_I=c(t)|\frac{\hbar\omega}{2}\rangle_s|0\rangle_R+\sum_{\mu=1}^3\int
d^3q
D_{q\mu}(t)|-\frac{\hbar\omega}{2}\rangle_s|\vec{q},\mu\rangle_R
\end{equation}
where $|0\rangle_R$ and $|\vec{q},\mu\rangle_R$ are the vacuum state
and an exited state of the absorbing environment, respectively. The
coefficients $c(t)$ and $D_{\vec{q}\mu}(t)$ should be specified by
the Sch\"{o}dinger equation in interaction picture as
\begin{equation}\label {s17}
H'_I(t)|\psi(t)\rangle_I=i\hbar\frac{\partial}{\partial
t}|\psi(t)\rangle_I
\end{equation}
for initial condition $c(0)=1 , D_{\vec{q}\mu}(0)=0$.  Substituting
$ |\psi(t)\rangle_I$ from (\ref{s16}) into (\ref{s17}) and applying
$H'_I(t)$ into (\ref{s15}) leads to the following coupled
differential equations for the coefficients  $c(t)$ and
$D_{\vec{q}\mu}(t)$
\begin{equation}\label {s18}
i\hbar\dot{c}(t)=-\hbar \sum_{\lambda=1}^3\int d^3k
f(\omega_{\vec{k}})e^{i(\omega_0-\omega_{\vec{k}})t}(\frac{e_{\vec{k}\lambda
x}}{2}+\frac{e_{\vec{k}\lambda y}}{2i})D_{\vec{k\lambda}}(t)
\end{equation}
\begin{equation}\label {s18.1}
i\hbar\dot{D}_{\vec{k}\lambda}(t)=-\hbar
f^\star(\omega_{\vec{k}})e^{i(\omega_{\vec{k}}-\omega_0)t}(
\frac{e_{\vec{k}\lambda x}}{2}-\frac{e_{\vec{k}\lambda y}}{2i}) c(t)
\end{equation}
Integrating equation (\ref{s18.1}), one can find the coefficient
$D_{\vec{k\lambda}}(t)$ in terms of $c(t)$ as
\begin{equation}\label {s19}
D_{\vec{k}\lambda}(t)=f^\star(\omega_{\vec{k}})\int_0^t dt'
e^{i(\omega_{\vec{k}}-\omega_0)t'}c(t')[\frac{ie_{\vec{k}\lambda
x}}{2}-\frac{e_{\vec{k}\lambda y}}{2}]
\end{equation}
where the initial condition $D_{\vec{k}\lambda}(0)=0$ has been used.
Now substituting $D_{\vec{k\lambda}}(t)$ from (\ref{s19}) in
(\ref{s18}) and using the completeness  relation
$\sum_{\lambda=1}^3e_{\vec{k}\lambda i}e_{\vec{k}\lambda
j}=\delta_{ij}$ gives  us the following integro- differential
equation for $c(t)$
\begin{equation}\label {s20}
\dot{c}(t)=\int_0^t dt'\gamma(t-t')c(t')
\end{equation}
where
\begin{equation}\label {s20.1}
\gamma(t-t')=-\frac{1}{2}\int
d^3k|f(\omega_{\vec{k}})|^2e^{i(\omega_0-\omega_{\vec{k}})(t-t')}
\end{equation}
Here we restrict our attention to the weak-coupling regime where the
Markov approximation applies and replace $c(t')$ in integrand
(\ref{s20}) by $c(t)$. Also we estimate the integral $\int_0^t
dt'e^{i(\omega_0-\omega_{\vec{k}})(t-t')}$ for sufficiently  long
time as\cite{33}
\begin{equation}\label {s21}
\int_0^t dt'e^{i(\omega_0-\omega_{\vec{k}})(t-t')}\simeq
iP\frac{1}{\omega_0-\omega_{\vec{k}}}+\pi\delta(\omega_0-\omega_{\vec{k}})
\end{equation}
where $P$ denotes the cauchy principal  value. Finally combination
of the relations  (\ref{s20}), (\ref{s20.1}) and (\ref{s21})leads to
\begin{equation}\label {s22}
\dot{c}(t)=-(\beta+i\Delta)c(t)
\end{equation}
where
\begin{equation}\label {s22}
\beta=\frac{2\pi^2}{c^3}\omega_0^2
|f(\omega_0)|^2=\frac{\hbar}{2}Im[\tilde{\chi}(\omega_0)]
\end{equation}
is the decay rate of spontaneous emission of an initially excited
spin-$\frac{1}{2}$ and
\begin{equation}\label {s23}
\Delta=\frac{2\pi}{c^3} P\int_0^\infty
d\omega'\frac{\omega'^2|f(\omega')|^2}{\omega_0-\omega'}=\frac{\hbar}{2\pi}\int_0^\infty
d\omega'\frac{Im[\tilde{\chi}(\omega')]}{\omega_0-\omega'}
\end{equation}
is the shift frequency.
\section{Transition probabilities}
In this section we find some transition probabilities between eigen
states of  $\omega_0 S_z$, when the absorbing environment is in
thermal state $\rho^T_B=\frac{e^\frac{-H_B}{KT}}{Tr_B(
e^{\frac{-H_B}{KT}})} $,where $K$ is Boltzmann constant and $Tr_B$
denotes tracing over the  degrees of freedom of the environment. In
order to obtain transition probabilities, we find the density
operator of the total system in interaction picture using the
perturbation theory. The time evolution of the density operator of
the total system in interaction picture can be obtained as
\begin{equation}\label {s24}
\rho_I(t)=U^\dag_I(t)\rho_I(0)U_I(t)
\end{equation}
where $U_I(t)$ is the time-evolution operator in interaction picture
and up to the first order perturbation is as
\begin{equation}\label {s25}
U_I(t)=1-\frac{i}{\hbar}\int_0^t dt_1 H'_I(t_1)
\end{equation}
Substituting $H'_I(t)$ from (\ref{s15}) in (\ref{s25}) leads to
\begin{eqnarray}\label {s26}
&&U_I(t)=\nonumber\\
&&1-\frac{i}{\hbar}\left[-\sum_{\lambda=1}^3\int d^3k
f(\omega_{\vec{k}})(\frac{e_{\vec{k}\lambda
x}}{2}+\frac{e_{\vec{k}\lambda
y}}{2i})e^{\frac{i(\omega_0-\omega_{\vec{k}})t}{2}}\
\frac{\sin\frac{(\omega_0-\omega_{\vec{k}})t}{2}}{\frac{(\omega_0-\omega_{\vec{k}})}{2}}
a_{\vec{k}\lambda}(0)S_+(0)\right]\nonumber\\
&&+\frac{i}{\hbar}\left[\sum_{\lambda=1}^3\int d^3k
f^\star(\omega_{\vec{k}})(\frac{e_{\vec{k}\lambda
x}}{2}-\frac{e_{\vec{k}\lambda
y}}{2i})e^{\frac{i(\omega_0-\omega_{\vec{k}})t}{2}}\frac{\sin\frac{(\omega_0-\omega_{\vec{k}})t}{2}}{\frac{(\omega_0-\omega_{\vec{k}})}{2}}
a^\dag_{\vec{k}\lambda}(0)S_-(0)\right]
\end{eqnarray}
In order to compute the transition probabilities between eiggen
states of $\omega_0S_z$ we need the reduced density operator of the
system which is defined by $ \rho_{sI}(t)=Tr_B[\rho_I(t)]$. Let the
initial density operator of the total system in interaction picture
at $t=0$ is $ \rho_I(0)=|\frac{\hbar\omega_0}{2}\rangle\langle
\frac{\hbar\omega_0}{2}|\otimes\rho^T_B$, then the reduced density
operator is obtained as
\begin{eqnarray}\label {s27}
&&\rho_{sI}(t)=|\frac{\hbar\omega_0}{2}\rangle\langle
\frac{\hbar\omega_0}{2}|\nonumber\\
&& +\frac{1}{2}\int d^3k|f(\omega_{\vec{k}})|^2
\frac{\frac{\sin^2(\omega_{\vec{k}}-\omega_0)t}{2}}{(\frac{\omega_0-\omega_{\vec{k}}}{2})^2}\frac{e^{\frac{\hbar
\omega_{\vec{k}}}{K T}}} {e^{\frac{\hbar \omega_{\vec{k}}}{K
T}}-1}|\frac{-\hbar\omega_0}{2}\rangle\langle
\frac{-\hbar\omega_0}{2}|
\end{eqnarray}
where the completeness relation $\sum_{\lambda=1}^3
e_{\vec{k}\lambda i}e_{\vec{k}\lambda j}=\delta_{ij}$ and
\begin{equation}
Tr_B[a^\dag_{\vec{k}\lambda}\rho^T_B
a_{\vec{k'}\lambda'}]=\delta(\vec{k}-\vec{k}')\delta_{\lambda\lambda'}
\frac{e^{\frac{\hbar\omega_{\vec{k}}}{KT}}}{\frac{\hbar\omega_{\vec{k}}}{KT}-1}
\end{equation}
have been used. Now if a measurement is done on the Hamiltonian
$\omega_0S_z$ the probability that the eiggenvalue
$-\frac{\hbar\omega_0}{2}$ is obtained for sufficiently long times
is
\begin{equation}\label{s28}
\Gamma_{|\frac{\hbar\omega_0}{2}\rangle\rightarrow
|-\frac{\hbar\omega_0}{2}\rangle}=Tr_s\left[|-\frac{\hbar\omega_0}{2}\rangle\langle-\frac{\hbar\omega_0}{2}|\rho_{sI(t)}\right]=\hbar
tIm[\tilde{\chi}(\omega_0)] \frac{e^{\frac{\hbar \omega_0}{K T}}}
{e^{\frac{\hbar \omega_0}{K T}}-1}
\end{equation}
where $Tr_s$ denotes tracing over the degrees of freedom of
spin-$\frac{1}{2}$ particle.\\
Now assume the initial density operator of the total system in
interaction picture is $
\rho_I(0)=|\frac{-\hbar\omega_0}{2}\rangle\langle
\frac{-\hbar\omega_0}{2}|\otimes\rho^T_B$. It is easy to show that
the reduced density operator of system is
\begin{eqnarray}\label {s29}
&&\rho_{sI}(t)=|\frac{-\hbar\omega_0}{2}\rangle\langle
\frac{-\hbar\omega_0}{2}| +\frac{1}{2}\int
d^3k|f(\omega_{\vec{k}})|^2
\frac{\sin^2\frac{(\omega_{\vec{k}}-\omega_0)t}{2}}{(\frac{\omega_0-\omega_{\vec{k}}}{2})^2}\frac{1}
{e^{\frac{\hbar \omega_{\vec{k}}}{K
T}}-1}|\frac{\hbar\omega_0}{2}\rangle\langle
\frac{\hbar\omega_0}{2}|\nonumber\\
&&
\end{eqnarray}
where
\begin{equation}
Tr_B[a_{\vec{k}\lambda}\rho^T_B
a^\dag_{\vec{k'}\lambda'}]=\delta(\vec{k}-\vec{k}')\delta_{\lambda\lambda'}
\frac{1}{\frac{\hbar\omega_{\vec{k}}}{KT}-1}
\end{equation}
has been used.  A simple calculation shows that in this case if the
Hamiltonian$\omega_0S_z$ is measured the probability that the
eiggenvalue $\frac{\hbar\omega_0}{2}$ is obtained for very long
times is
\begin{equation}\label{s30}
\Gamma_{|-\frac{\hbar\omega_0}{2}\rangle\rightarrow
|\frac{\hbar\omega_0}{2}\rangle}=Tr_s\left[|\frac{\hbar\omega_0}{2}\rangle\langle\frac{\hbar\omega_0}{2}|\rho_{sI(t)}\right]=\hbar
t I m[\tilde{\chi}(\omega_0)] \frac{1} {e^{\frac{\hbar \omega_0}{K
T}}-1}
\end{equation}
From this transition probability it is clear that at very low
temperature $\Gamma_{|-\frac{\hbar\omega_0}{2}\rangle\rightarrow
|\frac{\hbar\omega_0}{2}\rangle}$ tends to zero.
\section{Summary and conclusion }
The absorbing environment of a spin-$\frac{1}{2}$ particle is
modeled by a continuum set of harmonic oscillators. A coupling
function is introduced that couple the spin and the environment. A
susceptibility function is attributed to the environment that is
obtained in terms of the coupling function. It was shown that the
susceptibility function satisfies the Kramers-Kronig relations as in
electrodynamics. The quantum  Langevin -Schr\"{o}dinger equation is
obtained as equation of motion of the spin in Heisenberg picture. It
is assumed that the spin is in a uniform external  magnetic field
along the $z$ axis and spontaneous emission of the spin is
calculated in the presence of the absorbing environment. The
spontaneous emission and the shift frequency of the spin is obtained
in terms of the imaginary part of the susceptibility function of the
environment in frequency domain. Some transition probabilities
between eigen states of the spin ( embedded  in the uniform magnetic
field) is computed when the environment is in thermal state. It is
shown when the temperature of the environment tends to zero there
exist an irreversible current of energy from the spin to the
environment.

\end{document}